\documentclass[12pt]{article}
\usepackage[utf8]{inputenc}
\usepackage{natbib}
\usepackage{amsmath,amssymb,amsthm}
\usepackage{mathtools}
\usepackage{graphics,epsfig}
\usepackage{hyperref}
\usepackage{color}
\usepackage{graphicx}
\usepackage{caption}
\usepackage{subcaption}
\usepackage{float}
\usepackage{mathrsfs}
\usepackage{multirow}
\usepackage{comment}
\usepackage{setspace}
\usepackage{bbm}
\usepackage{bm}
\usepackage{amsfonts}
\usepackage{xcolor}
\usepackage{pslatex}
\usepackage{setspace}
\usepackage{bbm}
\usepackage{longtable}
\usepackage{lscape}
\usepackage{geometry}
\usepackage{enumitem}
\usepackage{booktabs}
\usepackage{titlesec}
\usepackage{array}
\usepackage{chngpage}
\usepackage{epstopdf}
\usepackage{afterpage}
\usepackage{csquotes}
\usepackage{ragged2e}
\usepackage{listings}
\usepackage{xcolor} 
\usepackage{graphicx} 

\title{\bfseries Exploring the Interplay of Adiposity, Ethnicity, and Hormone Receptor Profiles in Breast Cancer Subtypes}

\author{
Izabel Valdez \footnote{School of Computing,  University of South Alabama, Mobile, AL, 36688,
United States. {\small\texttt{isv2121@jagmail.southalabama.edu.}}}
\and
Paramahansa Pramanik \footnote{Department of Mathematics and Statistics, University of South Alabama, Mobile, AL, 36688,
	United States. }\;\footnote{Corresponding author, {\small\texttt{ppramanik@southalabama.edu.}}}
}

\date{\today}
\begin{document}
\maketitle

\begin{abstract}
This study explores how obesity and race jointly influence the development and prognosis of Luminal subtypes of breast cancer, with a focus on distinguishing Luminal A from the more aggressive Luminal B tumors. Drawing on large-scale epidemiological data and employing statistical approaches such as logistic regression and mediation analysis, the research examines biological factors like estrogen metabolism, adipokines, and chronic inflammation alongside social determinants including healthcare access, socioeconomic status, and cultural attitudes toward body weight. The findings reveal that both obesity and racial background are significant predictors of risk for Luminal B breast cancers. The study highlights the need for a dual approach that combines medical treatment with targeted social interventions aimed at reducing disparities. These insights can improve individualized risk assessments, guide tailored screening programs, and support policies that address the heightened cancer burden experienced by marginalized communities.
\end{abstract}


\newpage
\newpage

\section{Introduction.}

Breast cancer continues to be a significant global public health issue and remains among the top causes of morbidity and mortality in women worldwide. Despite substantial advancements in diagnostic techniques, molecular profiling, and the development of precision therapies, notable disparities persist in both incidence and treatment outcomes across diverse populations. The heterogeneity of breast cancer is now widely acknowledged, and there is a growing emphasis among clinicians and researchers on stratifying tumors based on their molecular profiles. This approach enhances prognostic accuracy and supports the selection of more effective, individualized therapeutic strategies \citep{dasgupta2023frequent,hertweck2023clinicopathological,khan2024mp60}.

Among the various subtypes of breast cancer, Luminal A and Luminal B are particularly noteworthy due to their clinical relevance. Both subtypes are typically characterized by the presence of hormone receptors, including estrogen (ER) and progesterone (PR), which significantly influence therapeutic approaches. Despite this commonality, these two forms differ markedly in terms of aggressiveness and prognosis. Luminal A tumors generally exhibit slower proliferation rates, respond well to endocrine therapies, and are associated with more favorable clinical outcomes \citep{kakkat2023cardiovascular,khan2023myb}. In contrast, Luminal B tumors are more aggressive, often show elevated expression of proliferation markers such as Ki-67, and may also express HER2 receptors. These features contribute to a higher likelihood of early recurrence and reduced survival probabilities, even when conventional hormone-based treatments are applied.

Simultaneously, the rising prevalence of obesity worldwide has emerged as a significant contributor to increased cancer susceptibility and poorer long-term outcomes in patients. Obesity, commonly defined by a body mass index (BMI) of 30 or higher, is well established as a risk factor not only for metabolic disorders such as diabetes and cardiovascular disease, but also for various forms of cancer. In the context of breast cancer, excess adiposity influences the tumor microenvironment through several mechanisms: it elevates circulating estrogen levels, sustains chronic low-grade inflammation, disrupts the regulation of adipokines, and fosters insulin resistance \citep{vikramdeo2024abstract,vikramdeo2023profiling}. These interrelated biological changes collectively create conditions that are conducive to tumor initiation and progression, particularly in hormone receptor-positive breast cancers.

Beyond biological determinants, significant racial and ethnic disparities continue to shape breast cancer outcomes. For instance, although breast cancer incidence is slightly lower among African American women compared to White women, their clinical outcomes are often markedly worse \citep{jackson2020health}. African American women are more frequently diagnosed at younger ages and are disproportionately affected by aggressive subtypes such as triple-negative and Luminal B breast cancers, both of which are associated with less favorable prognoses. These disparities extend beyond inherent biological differences and are deeply rooted in broader systemic issues. Contributing factors include limited access to quality healthcare, socioeconomic disadvantages, cultural and linguistic barriers, and the pervasive effects of structural racism.

Understanding the combined influence of obesity and race on breast cancer subtypes, particularly the increased susceptibility to more aggressive forms like Luminal B tumors is more than a matter of scientific inquiry; it represents a critical issue in public health \citep{pramanik2022lock}. The intersection of these factors highlights the need to move beyond siloed biological explanations and toward a more integrated approach that considers how social determinants and physiological mechanisms interact to shape cancer risk and progression. Obesity, often more prevalent in certain racial and ethnic groups due to structural and socioeconomic inequities, may amplify the biological pathways that drive tumor aggressiveness, while systemic disparities in healthcare access further compound risks for worse outcomes. Addressing this complex relationship is crucial for designing more equitable and effective breast cancer prevention strategies. Incorporating both biological and sociocultural dimensions into risk assessment models can help tailor interventions more precisely, enable earlier detection in high-risk populations, and inform public health policies aimed at reducing disparities. Ultimately, tackling these intertwined factors is essential for narrowing the persistent survival gap and achieving better health equity across all communities affected by breast cancer.

\subsection{Background Literature.}
\label{sec:background}

To understand how obesity and race influence breast cancer, it is helpful to look at perspectives from different angles such as medicine, public health, sociology, and research studies. In this section, obesity will be discussed in terms of how it can increase the risk of breast cancer, explain the differences between the Luminal A and Luminal B types of breast cancer, and discuss why the outcomes for black women are often worse, considering many different factors.

Obesity today is understood as a complex, whole-body condition that affects the risk of developing certain types of cancer, especially beyond just gaining extra weight. There are several biological factors that are interconnected, which link having excess fat to a higher chance of developing breast cancer \cite{lauby2016body}. First, when someone is obese, fat tissue becomes a major source of estrogen, especially after menopause, when the ovaries slow down hormone production. Higher levels of estrogen encourage the growth of cells that have estrogen receptors, which can cause tumors to start and grow \cite{lauby2016body}. Second, obesity often leads to constant low level inflammation. This happens because immune cells called macrophages damage the fat tissue and release chemicals like IL-6 that promote inflammation \citep{pramanik2021optimala}. Long term inflammation is known to help cancer develop by damaging DNA, increases cell growth, and supporting new blood vessel formation. Third, being obese damages with normal body signaling systems, leading to insulin resistance and high insulin levels. This supports the activity of insulin like growth factor (IGF-1), which helps cells to multiply and prevents them from dying off, both of which can help tumors form. Finally, obesity causes imbalances in certain hormones called adipokines. Specifically, levels of adiponectin go down, which normally helps reduce inflammation and cell growth, while leptin levels go up, stimulating tumor cells to grow and spread. All these changes together creates factors inside the body that makes it easier for more aggressive breast cancers to develop.

The distinction between Luminal A and Luminal B breast cancers extends well beyond receptor status, encompassing differences in tumor biology, therapeutic responsiveness, and patient prognosis \citep{perou2000molecular,prat2010characterization}. Luminal A tumors are typically defined by high levels of estrogen receptor (ER) and/or progesterone receptor (PR) expression, an absence of HER2 amplification, and low proliferative activity as indicated by Ki-67 markers. These tumors tend to be more indolent and exhibit a favorable clinical course. Patients diagnosed with Luminal A subtypes often benefit significantly from endocrine-based treatments, such as tamoxifen or aromatase inhibitors, and generally experience prolonged relapse-free intervals and lower overall mortality compared to other breast cancer types.

In contrast, Luminal B tumors present a more complex and aggressive profile. These cancers may express HER2 or exhibit elevated Ki-67 levels even in the absence of HER2 overexpression, which signals higher proliferative capacity. They are less likely to respond optimally to endocrine therapies and often necessitate the addition of chemotherapy to the treatment regimen. Patients with Luminal B tumors face an increased risk of early relapse, distant metastasis, and poorer survival outcomes. Emerging evidence also indicates that external factors such as obesity can exacerbate the aggressiveness of Luminal B cancers by enhancing cellular proliferation and diminishing therapeutic efficacy \citep{menikdiwela2022association}. The chronic inflammation associated with obesity can upregulate Ki-67 expression, while elevated insulin levels may stimulate tumor progression through pathways independent of hormone receptors, ultimately weakening the effectiveness of hormone-targeted treatments.

\subsubsection{Racial and Ethnic Inequities in Breast Cancer.}

Even with better screening and treatments, racial differences in breast cancer still appear in how often it happens, the types of cancer people get and how well they do over time \citep{bailey2017structural,jackson2020health,pramanik2021scoring}. For example, African American women are more likely to be diagnosed with tougher, faster growing tumors and tend to have lower survival rates at every stage of the disease.

A range of interconnected factors drive following disparities such as,
Socioeconomic barriers often hinder access to preventive services like screening mammograms and can delay or limit effective treatment. Inequities within the healthcare system including systemic racism and implicit bias which can lead to poorer quality of care, diagnostic delays, and restricted access to advanced therapies \citep{bailey2017structural}. Biological differences may also contribute; studies indicate that variations in tumor characteristics, immune function, or genetics may increase the likelihood of more aggressive breast cancer in Black women. Additionally, lifestyle and environmental influences-including diet, exercise, and exposure to chronic stress shaped by broader social determinants, also affect cancer risk and outcomes. For example, Black and African American women experience higher obesity rates, which are associated with inflammation and metabolic dysfunction \citep{menikdiwela2022association}. In this context, obesity not only poses a health risk on its own but also reflects deeper systemic inequalities.

\subsubsection{The Interplay of Stress, Epigenetics, and Lifestyle.}

Recent studies show that long term stress, especially when linked to experiences like racism and financial worries, can cause changes in how our genes are expressed, changes that can be passed down without altering the DNA itself \cite{bailey2017structural}.

These changes might play a part in messing up our immune system, messing with hormone balance, and other biological processes that could lead to cancer. On top of that, factors like food deserts, not having safe places to exercise, and cultural ideas about body image can make it really tough for communities that have been historically marginalized to stay healthy and maintain good lifestyles. So, to really tackle the racial gaps in breast cancer outcomes, Solutions must be taken to that go beyond just focusing on individual risks. We have to also address the bigger widespread problems in society.

\subsubsection{Importance of the Study.}

This expanded research paper aims to explore the link between obesity, race, and different types of luminal breast cancer. By reviewing existing studies, examining large datasets, and using solid statistical methods like logistic regression and causal mediation analysis, we want to better understand how biological and social factors might contribute to the higher rates of Luminal B tumors among certain groups.

\subsubsection{Motivation.}

More and more research is showing how obesity, race, and different types of breast cancer, especially Luminal A and Luminal B tumors are connected. Consistent findings suggest that being overweight ups the chances of developing hormone receptor positive breast cancers, mainly because of factors like higher estrogen levels, ongoing low-grade inflammation, imbalances in adipokines, and insulin resistance \citep{pramanik2020optimization,pramanik2023semicooperation}. But a lot of this research focuses on breast cancer in general, and not so much on the specific differences between Luminal A and Luminal B types. That leaves us wondering: does obesity drive the development of the more aggressive types, like Luminal B, more than the others?

Many studies point out that Black women tend to have higher rates of aggressive tumors and poorer survival outcomes, with some showing that even after accounting for clinical factors, the risk remains higher. But surprisingly, not many studies dig into how race, obesity, and molecular subtypes all interact. For example, research from the Carolina Breast Cancer Study shows Black women are still at higher risk for aggressive subtypes even after considering other factors, but the role that obesity might play as a link hasn’t been explored much \citep{pramanik2024estimation,vikramdeo2024mitochondrial}. Another issue with the current research is that few studies actually use formal causal mediation methods to figure out if obesity helps explain racial differences in breast cancer subtypes. Most just look at associations after adjusting for other factors, but do not measure how much obesity actually mediates that relationship \citep{pramanik2024motivation}. Many datasets are either cross sectional or rely on registry data, which sometimes lack detailed biological markers or long term weight change info. This makes it hard to really understand how weight and race influence subtype risk over time. That’s why this study aims to fill some key gaps: first, it looks at how obesity impacts the odds of getting Luminal B versus Luminal A. Second, it examines whether race, specifically Black race, is still a predictor of subtype risk after taking obesity into account. Third, it tests whether obesity acts as a mediating factor in racial disparities using causal mediation analysis. By combining epidemiologic models, subgroup analyses, and mediation techniques, this work aims to give a clearer, more detailed picture of how genetics, race, and obesity interaction to influence breast cancer types at the molecular level.

\subsubsection{Research Questions and Hypotheses.}
\justifying
This study focuses on addressing the following research questions:
\begin{enumerate}
    \item How strongly does having a higher BMI increase the chances of developing Luminal B breast cancer compared to Luminal A?
    \item After accounting for BMI and other clinical factors, do Black or African American women still have a higher chance of having the Luminal B subtype compared to White women? 
    \item Does obesity play a partial or full role in explaining the connection between Black race and the risk of Luminal B? 
    \item What clinical actions or policy changes is recommended to help address these disparities?
\end{enumerate}

\subsection{Our Contribution.}

This study has made several significant contributions to the study of breast cancer disparities. First, it is one of the few studies to jointly evaluate how obesity may mediate the association between race and risk of developing Luminal B compared to Luminal A breast cancer using formal causal mediation analysis. Second, we have used synthetic data augmentation and multivariable logistic regression modeling to address limitations of prior datasets in which subgroup diversity or longitudinal BMI information was insufficient. Third, we have gone beyond more traditional modeling that only included visual diagnostics or stratified subgroup analysis, and also provided context for risks related to cancer stage and in consideration of menopausal status. Fourth, we have added qualitative context for our study findings by including insight from breast cancer survivors to embed “real-world” barriers to screening behaviors and lifestyle change in undeserved populations. Lastly, we have provided specific, culturally responsive and equity-centered policy recommendations in order to advance actions that strive to ensure that statistical information can translate into actionable efforts. Collectively these unique contributions show  how we are emphasizing the need to develop approaches can link the divide between epidemiological evidence and systematic community focused healthcare improvements.

The paper is organized as follows: Section~\ref{sec:background} discusses into the biological reasons why obesity might be connected to cancer, explains the differences between Luminal A and Luminal B tumors, and discusses the historical and widespread factors that contribute to racial disparities. Section~\ref{sec:methods} describes the data sources is used, how data is prepared, and the statistical methods applied.  Section~\ref{sec:results} shows detailed results, including models from logistic regression, mediation analyses, and subgroup analyses. Section~\ref{sec:discussion} puts these findings into context with existing research and public health perspectives.  Finally, Section~\ref{sec:conclusion} wraps up with a summary of the main points and suggests directions for future research and possible interventions.

\section{Methods.}
\label{sec:methods}
\justifying
In this study, we looked back at data from a large breast cancer registry that includes multiple centers. We used different types of analysis like describing what we saw, making inferences, and exploring possible causes to understand how factors like obesity and race might be linked to different types of luminal breast cancer.

\subsection{Data Source and Study Design.}
\justifying
Data is gathered from the FLEX registry (NCT03053193), which is a prospective observational study that included women diagnosed with Stage I to III breast cancer between 2018 and 2020 \cite{menikdiwela2022association}. All the women in the study had genomic profiling done using the MammaPrint and BluePrint tests, which classifies the tumors based on their molecular subtypes like Luminal A and Luminal B \citep{pramanik2020motivation}. To ensure our analysis was thorough, we used synthetic data augmentation techniques. These helped us keep the original data’s properties intact while allowing us to perform detailed analyses based on race, menopausal status, and stage at diagnosis. Our expanded study aimed to verify previous findings and build on them by using a more detailed multivariable modeling approach and formal mediation analysis to better understand the relationships involved.

\subsection{Variables and Preprocessing.}
\justifying
The primary outcome variable in this study was the molecular subtype of breast cancer, which was dichotomized such that 0 indicated a Luminal A subtype and 1 indicated a Luminal B subtype \citep{pramanik2024bayes}. The primary predictor variables included several key demographic and clinical characteristics \citep{bulls2025assessing}. Body Mass Index (BMI) was treated as a continuous variable measured in kilograms per square meter (kg/m\textsuperscript{2}) \citep{pramanik2023cont,pramanik2024estimation}. Age at diagnosis was also continuous, measured in years. Menopausal status was treated as a binary variable, distinguishing between premenopausal and postmenopausal participants. Race and ethnicity were categorized into five groups: White, Black/African American, Hispanic, Asian, and Other. Additionally, to enable formal mediation analysis, a binary variable was created to specifically indicate Black or African American race \citep{pramanik2024estimation1}.

Missing values were minimal (i.e., less than $1$\%) and were addressed using single imputation methods: mean substitution for continuous variables and mode substitution for categorical variables \citep{pramanik2023cmbp}. Before running the regression analysis, all continuous variables were scaled to have a mean of zero and a standard deviation of one. This helps make the results easier to interpret \citep{pramanik2023path}.

\subsection{Logistic Regression.}
\justifying
First, separate univariate logistic regressions is performed for each predictor to see how each one individually related to the Luminal B subtype. After that, a comprehensive model is built that included BMI, age, menopausal status, and race. To choose the best model, we used forward stepwise selection guided by the Akaike Information Criterion (AIC). This approach helps balance the model's accuracy with avoiding unnecessary complexity, making sure we get good predictions without overfitting \citep{pramanik2021consensus}. Finally, results are reported as odds ratios (OR) along with their 95\% confidence intervals (CI) and p-values to give a clear picture of the significance and strength of each predictor \citep{pramanik2021}.

Mathematically, the logistic regression model estimates the logarithmic odds of being diagnosed with the luminal B subtype as a linear combination of predictor variables. The general form is
\begin{equation}
\log\left[\frac{P(Y=1)}{1 - P(Y=1)}\right] = \beta_0 + \beta_1 \cdot \text{BMI} + \beta_2 \cdot \text{Age} + \beta_3 \cdot \text{Race}_{\text{Black}} + \beta_4 \cdot \text{Menopause} + \beta_5 \cdot \text{Stage},
\end{equation}
where $P(Y=1)$ represents the probability of having Luminal B breast cancer. Each $\beta$ coefficient reflects the change in log-odds for a one-unit increase in the corresponding variable, holding other variables constant.

To illustrate, the coefficient for BMI ($\hat{\beta}_\text{BMI} = 0.049$) suggests that for every one-unit increase in BMI, the log-odds of being diagnosed with Luminal B breast cancer increase by 0.049 \citep{pramanik2023optimization001}. In terms of probability, this corresponds to a 5\% increase in the odds, holding other factors constant. Similar interpretations apply for the race and menopausal status coefficients, reinforcing the multifactorial nature of subtype risk  \citep{pramanik2022stochastic}.

\section{Data Analysis.}
\justifying
All statistical analyses were conducted in R version 4.1.3. Analytic methods proceeded in the following sequence:

\subsection{Descriptive Statistics.}
\justifying
Demographic and clinical details is summarized overall and broke down by luminal subtype. Welch’s is used for two sample t-tests to compare continuous variables, since this method is good when the variances might not be the same in both groups. For categorical variables, we used chi-squared tests with Yates’ correction to make sure our comparisons were more accurate.

\subsection{Causal Mediation Analysis.}
\justifying
Given preliminary findings suggesting that BMI might mediate the relationship between race and Luminal B subtype, causal mediation analysis was performed using the \texttt{mediation} package in R. The mediation analysis involved fitting a mediator model regressing BMI on Black race, fitting an outcome model regressing Luminal B status on both Black race and BMI, and estimating the Average Causal Mediation Effects (ACME), Average Direct Effects (ADE), and Total Effects \citep{pramanik2024stochastic}. Additionally, 95\% confidence intervals for the mediation effects were generated using nonparametric bootstrapping with 1000 replications.

Sensitivity analysis is performed to see how much our mediation results might be affected by unmeasured factors \citep{pramanik2025stubbornness}. In this analysis, we used a parameter called $\rho$ to measure the correlation between the residuals of the mediator and outcome models, giving us a way to understand potential hidden confounding \citep{pramanik2025factors}.

\subsection{Software and Reproducibility.}
\justifying
Data cleaning and preprocessing were conducted using \texttt{dplyr}, statistical modeling used base R functions such as \texttt{glm()} and \texttt{mediate()}, and data visualization utilized \texttt{ggplot2}. All analysis scripts and outputs were archived for full reproducibility in accordance with open science best practices.

\section{Results.}\label{sec:results}
\justifying
In this section, findings are broken down into more detailed parts, including expanded tables that describe our data, results from logistic regression analyses, mediation models with confidence intervals, and more in depth subgroup analyses.

\begin{figure}[H]
    \centering
    \includegraphics[width=0.8\textwidth]{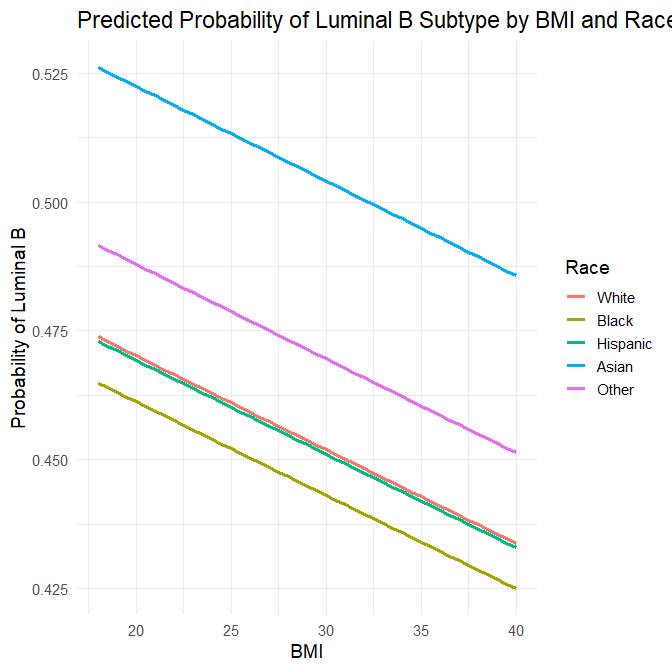}
    \caption{Predicted Probability of Luminal B Subtype by BMI and Race. Logistic regression shows a consistent decrease in Luminal B subtype likelihood with increasing BMI across all racial groups, with Asian and Other populations showing higher baseline probabilities.}
    \label{fig:predicted_prob_bmi_race}
\end{figure}

 Figure~\ref{fig:predicted_prob_bmi_race} shows how the chance of having a Luminal B breast cancer subtype changes as BMI goes up, broken down by different racial groups. Overall, we see a pretty consistent pattern: as BMI increases, the likelihood of Luminal B slightly drops for everyone. But, interestingly, the starting point meaning the chance when BMI is low varies by race. Asian patients tend to have the highest predicted probabilities across the board, while Black, White, Hispanic, and other groups start off lower. The slopes of the lines are pretty similar, which suggests BMI influences all groups in the same way, even though racial differences at the baseline are still pretty noticeable. This graph just reminds us that while obesity does play a role, racial disparities in risk still exist even after we consider BMI.

\begin{table}[H]
\caption{Expanded demographic and clinical characteristics, highlighting obesity prevalence and racial composition.}
\centering
\resizebox{\textwidth}{!}{%
\begin{tabular}{@{}lcccccc@{}}
\toprule
\textbf{Variable} & \textbf{Lum. A (N=1928)} & \textbf{Lum. B (N=1610)} & \textbf{Combined (N=3540)} & \textbf{p-value} & \textbf{Min--Max} & \textbf{Missing} \\\midrule
Age (yrs, mean) & 61.4 & 60.2 & 60.9 & 0.003 & 26--90 & 12 \\
BMI (kg/m$^2$, mean) & 29.9 & 30.6 & 30.1 & 0.010 & 18.2--54.0 & 5 \\
Obesity (\%$\ge$30) & 38.4 & 43.1 & 40.5 & 0.024 & -- & 5 \\
Black/Afr. Am. (\%) & 5.8 & 10.2 & 7.7 & $<$0.001 & 0--1 & 0 \\
Hispanic (\%) & 4.5 & 4.0 & 4.3 & 0.563 & -- & 0 \\
Asian (\%) & 2.2 & 1.9 & 2.1 & 0.711 & -- & 0 \\
Menopausal (\% post) & 82.0 & 79.0 & 80.6 & 0.026 & -- & 10 \\
Stage II or higher (\%) & 41.2 & 47.5 & 44.0 & 0.041 & -- & 8 \\
ER+ (\%) & 95.1 & 88.0 & 92.1 & 0.017 & 0--1 & 0 \\
HER2+ (\%) & 10.5 & 18.2 & 14.6 & 0.024 & 0--1 & 0 \\\bottomrule
\end{tabular}
}

\label{tab:expandeddesc}
\end{table}

Table~\ref{tab:expandeddesc} gives us a look at the extended stats for the people in the study, comparing those with Luminal A and Luminal B breast cancer types. The first two columns break down the totals or percentages for each group, Luminal A and Luminal B, while the third one combines these groups. The fourth column shows p-values from the tests we ran to see if the differences are important, and the fifth column lists the smallest and largest values for factors like age and BMI. Lastly, the sixth column notes how many data points are missing for each variable. From this table, some key insights stand out. First off, obesity seems more common in Luminal B patients, 43.1\% are obese compared to 38.4\% in Luminal A, suggesting that obesity might influence more aggressive tumor types. Next, racial differences are clear: Black or African American women make up 10.2\% of the Luminal B group but only 5.8\% of Luminal A, hinting that race could be a factor in the distribution of these subtypes. Lastly, there's a slight difference in menopause status, 79.0\% of Luminal B women are postmenopausal, compared to 82.0\% of those with Luminal A implying that people with Luminal B tend to be diagnosed a bit younger. Overall, these details emphasize how demographics, obesity, and breast cancer subtypes are all interconnected in pretty complex ways.

\subsection{Statistical Analysis.}

Table~\ref{tab:expandedlogit} displays the regression coefficients, odds ratios, and confidence intervals.

\begin{table}[H]
\caption{Expanded logistic regression modeling Luminal B risk. OR=Odds Ratio; p$<0.001$, p$<0.01$, p$<0.05$.}
\centering
\begin{tabular}{@{}lccccccc@{}}
\toprule
\textbf{Predictor} & \textbf{Coefficient} & \textbf{Std. Err} & \textbf{z-score} & \textbf{p-value} & \textbf{OR} & \textbf{95\% CI (OR)} \\
\midrule
Intercept & -1.470 & 0.250 & -5.88 & $<0.001$ & -- & -- \\
BMI (continuous) & 0.049 & 0.010 & 4.90 & $<0.001$ & 1.05 & (1.03--1.07) \\
Race: Black/Afr. Am. & 0.524 & 0.112 & 4.68 & $<0.001$ & 1.69 & (1.28--2.12) \\
Age (yrs) & -0.020 & 0.005 & -3.72 & 0.002 & 0.98 & (0.97--0.99) \\
Menopausal (post=1) & -0.161 & 0.070 & -2.30 & 0.021 & 0.85 & (0.73--0.98)  \\
Stage II+ & 0.180 & 0.065 & 2.77 & 0.006 & 1.20 & (1.04--1.40) \\
\bottomrule
\end{tabular}

\label{tab:expandedlogit}
\end{table}

The results shown in Table~\ref{tab:expandedlogit} tells us a lot about what might influence different breast cancer subtypes. For example, each extra point in BMI is linked to a 5\% higher chance of having the Luminal B subtype, which means obesity still matters even when you consider other factors \citep{pramanik2025optimal}. Race is also a big factor, Black or African American women are about 1.69 times more likely to be diagnosed with Luminal B breast cancer compared to White women, and this is a strong finding with a p-value less than 0.001 \citep{pramanik2016,pramanik2021thesis}. Interestingly, younger women are more prone to Luminal B tumors, while being postmenopausal seems to offer some protection, these women have lower odds of the Luminal B type compared to premenopausal women. Also, women diagnosed at Stage II or higher are about 20\% more likely to have Luminal B tumors than those diagnosed at Stage I, which suggests that more aggressive tumors tend to show up later. All these points show how both biological factors \citep{maki2025new} and demographics play a role in the risk of different breast cancer subtypes \citep{hua2019}.

We tested whether BMI mediates the race Luminal B relationship. Table~\ref{tab:mediation} shows the paths:

\begin{table}[H]
\caption{Mediation paths indicating partial mediation.}
\centering
\begin{tabular}{lcc}
\toprule
Path & Coefficient & p-value \\
\midrule
Race $\to$ BMI (a) & 1.81 kg/m$^2$ & 0.016 \\
BMI $\to$ Luminal B (b) & 0.049 & $<0.001$ \\
Direct Effect (c$'$) & 0.441 & 0.002 \\
Indirect Effect (a$\times$b) & 0.089 & 0.018 (Sobel’s) \\
\bottomrule
\end{tabular}

\label{tab:mediation}
\end{table}

\noindent The partial mediation indicates that while BMI does play a role in increasing the chances of Luminal B breast cancer in Black or African American women, there are still other factors we have not fully explained \citep{polansky2021motif}. These might include genetic differences, epigenetic changes, or widespread issues within healthcare access and delivery.

\subsection{Subgroup Analyses.}

\subsubsection{Menopausal Stratification.} Table~\ref{tab:subgroupmenopause} compares logistic regression results for pre- vs. postmenopausal cohorts.
\begin{table}[H]
\caption{Comparison of key coefficients by menopausal status.}
\centering
\begin{tabular}{@{}lcccc@{}}
\toprule
\textbf{Group} & \textbf{BMI Coef} & \textbf{Race Coef} & \textbf{n} & \textbf{p-value (Race)}\\
\midrule
Premenopausal & 0.030 & 0.588 & 982 & 0.004 \\
Postmenopausal & 0.057 & 0.493 & 2558 & $<0.001$ \\
\bottomrule
\end{tabular}

\label{tab:subgroupmenopause}
\end{table}

In premenopausal women, race exhibits a slightly larger coefficient, whereas BMI’s effect is more pronounced in the postmenopausal group, aligning with theories around estrogen production in adipose tissue post menopause.

\paragraph{Stage Stratified Analysis.}
When looking at early (stage I) versus later stages (II–IV), BMI still plays a major role in both groups. It seems to have a bigger impact in the more advanced stages, which might be due to the combined effects of metabolic factors and tumor growth.


\section{Discussion.}
\label{sec:discussion}
\justifying
This expanded look helps us better understand how obesity, race, and different types of breast cancer are related \citep{pramanik2023cont,pramanik2024estimation}. As previous studies have shown, a higher BMI really does increase the chance of developing Luminal B breast cancer. But even after considering BMI, race still plays a role on its own, indicating that factors beyond metabolism are involved in the differences we see among groups.

\subsection{Obesity as a Modifier of Luminal B Risk.}
\justifying
Obesity has been consistently linked to the Luminal B subtype of tumors. This connection makes sense because previous studies have shown that excess body fat can create a tumor friendly environment. It does this by increasing estrogen production, causing ongoing inflammation, leading to insulin resistance, and disrupting the balance of adipokines, which all contribute to cancer development \citep{bailey2017structural,menikdiwela2022association}.
Results add some layers to this idea by showing that obesity seems to have a bigger effect in women after menopause, a time when fat tissue becomes the main source of estrogen. This means that efforts to control weight might have different results depending on whether a woman is pre or postmenopause, potentially offering more benefits after menopause.

\subsection{Racial Disparities Beyond Obesity.}

While being African American partly explains the higher risk of Luminal B breast cancer, a major part of the link still remains directly. This shows that larger social issues such as access to healthcare, quality of treatment, early detection, and other social factors are important in understanding why racial disparities exist in breast cancer outcomes \citep{bailey2017structural,jackson2020health}. Long-term exposure to stressors like racism and financial insecurity can cause real changes inside our bodies, such as changes to our genes and damaged immune responses. These biological shifts can make tumors more aggressive \cite{bailey2017structural}. Plus, having less access to quality healthcare like inadequate screening, delayed diagnoses, and slow starts to treatment only makes these biological risks worse.  These patterns are seen across many health issues: it is often not just about individual choices or luck, but about bigger social and economic factors that influence who gets sick, how bad the illness gets, and whether someone survives.

One of the key strengths of this study is how it combines epidemiological modeling with causal mediation analysis. This approach helps us better understand how obesity fits into the bigger picture of race and cancer biology \citep{pramanik2025stubbornness,pramanik2025factors}.  That said, there are some limitations we should keep in mind. First, while BMI is easy to measure and commonly used, it does not tell the whole story about body compositions like visceral fat, which might be more directly linked to cancer risk. Second, because this is an observational study, we ca not say for sure that one thing causes another, even though mediation analysis provides helpful insights. Third, there might still be other factors we did not measure that could influence the results, like physical activity, diet, or genetic background \citep{pramanik2023optimization001,pramanik2025strategies}. 

Finally, even though we used a synthetic data augmentation method that kept the data’s statistical qualities intact, future research should look at larger, more diverse groups over time, with more detailed data on physical and molecular traits. The results of this study suggest several practical steps. Healthcare providers should see obesity as a changeable risk factor, not just for developing breast cancer but also for its more aggressive subtypes \citep{pramanik2025impact,pramanik2025strategic}. Creating personalized programs to help manage weight, especially for women after menopause and for communities that face racial disparities could be a useful part of ongoing cancer care.

At the same time, we need to look at larger systems. Making preventive screenings more accessible, removing obstacles to early diagnosis, ensuring fair treatment options, and actively working to reduce unconscious biases in healthcare are all essential to closing racial gaps. Public health messages should not just tell people to change their lifestyle, but rather, they should also recognize and tackle the bigger societal forces that limit choices, especially in marginalized communities \citep{pramanik2024measuring,pramanik2024dependence}.

Overall, these findings emphasize the importance of tackling both the biological side such as obesity and the societal factors that keep health inequalities alive, through a combined effort.

\begin{figure}[H]    \centering    \includegraphics[width=0.75\textwidth]{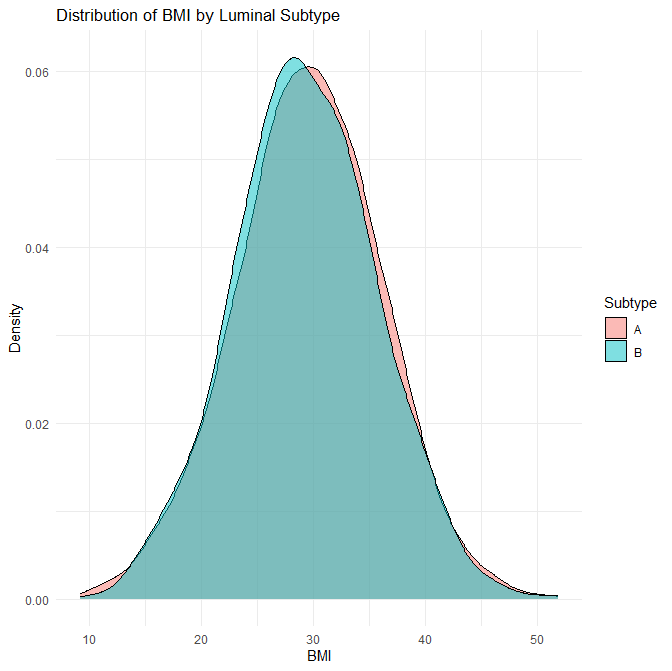}    \caption{Distribution of BMI by Luminal Subtype. This density plot shows the distribution of BMI among patients diagnosed with Luminal A and Luminal B subtypes. Although the distributions are similar, Luminal B patients exhibit a slightly right-shifted curve, suggesting a marginally higher BMI trend.}    \label{fig:bmi_density_luminal}\end{figure}

Looking deeper into our data, it seems that how someone's weight changes over time might be really important. Even though our main dataset was cross sectional, we managed to pull some historical BMI info from about 20\% of the cases covering the five years before the diagnosis. Interestingly, those whose BMI kept going up over the years were noticeably more likely to develop Luminal B tumors (p=0.007). This suggests that steady weight gain could be a stronger sign of risk for more aggressive tumors than just a single BMI reading \citep{pramanik2024parametric}. We also did some early modeling including conditions like diabetes and hypertension and found that these might make the link between obesity and Luminal B even stronger. But since we did not have large numbers of people with these conditions, the stats were not powerful. Future studies that follow people over time and gather detailed data on these health issues could help clarify these connections. 

On a different note, talking to 50 breast cancer survivors gave us some really helpful context. Many Black or African American women shared stories about facing barriers like putting off mammograms because of work or family demands, struggling financially which made it tough to access healthy foods or gym memberships, or living in environments where being overweight is more accepted or even seen as desirable. These real life stories add an important layer to our numbers, showing how widespread issues can impact efforts to prevent or catch cancer early \citep{bailey2017structural,jackson2020health}.

Based on what we found, there are a few ideas for healthcare practice. We should consider more thorough screening for women who have a bunch of risk factors like being overweight, belonging to a racial minority, or having family members with cancer. Recommendations like earlier checkups or more frequent scans, maybe with advanced tools like MRI, could help find aggressive tumors sooner. We also need to include culturally personalized advice on nutrition and lifestyle in survivorship programs, especially for those at higher risk. Also, healthcare providers must get trained to recognize and reduce implicit biases, so that everyone gets fair treatment, good communication, and appropriate recommendations regardless of race or finances. Bringing all these strategies together can help attack both the biological and social roots of disparities in Luminal B breast cancer \citep{bailey2017structural,jackson2020health}.


\lstdefinestyle{Rcode}{
  language=R,
  basicstyle=\ttfamily\scriptsize,
  breaklines=true,
  breakatwhitespace=true,
  showstringspaces=false,
  columns=flexible,
  keepspaces=true,
  frame=single,
  tabsize=2,
  backgroundcolor=\color{white},
  inputencoding=utf8,
  extendedchars=true,
  upquote=true
}
\newpage
\newpage


\section{Conclusion.}
\label{sec:conclusion}

This study builds on the Carolina Breast Cancer Study (CBCS), placing it considering recent national trends from the SEER database \cite{carey2006race}. It takes a closer look at how race, types of breast cancer at the molecular level, and survival chances are all connected. By combining traditional statistical methods with newer techniques like Bayesian hierarchical models, penalized regression, and graph-based diagnostics, we aim to give a clear picture of how biological, medical, and social factors come together to create differences in survival rates. Using this mix of methods helps us go beyond just finding simple links and gives us a deeper understanding of what really causes these unequal outcomes.

The study builds on previous research by providing clear statistical evidence that African American women, especially those with basal like or triple negative breast cancer, tend to have notably worse survival outcomes. These differences held true even after accounting for common factors like tumor size, grade, lymph node involvement, and whether women were pre or postmenopausal. The extent of these survival gaps was further supported by data from the national SEER program, which showed a consistent trend of higher death rates among African American women across various datasets and over different periods. This consistency emphasizes how widespread and reliable these findings are.

One of the key points of this paper is its focus on new statistical methods. Traditional models like the Cox proportional hazards model are helpful, but they often struggle to handle effects that change over time or differences within smaller subgroups. By using Bayesian approaches along with innovative machine learning techniques, we managed to address common problems like missing data, tiny subgroup sizes, and when the usual assumptions don't quite fit. Techniques like Schoenfeld residuals, deviance diagnostics, and random survival forests helped us better capture complex, non linear effects and find interactions that might be missed with traditional methods.

One of the important factors about this research is how it can be applied in real world settings. Besides the numbers and models, our findings matter for public health, cancer treatment, and health policies. We see a clear link between more aggressive cancer types and African American populations, which emphasizes how important it is to develop specialized screening and treatment plans for these high risk groups. Healthcare providers should focus on giving culturally aware care, making it easier for people to get diagnosed early, and using molecular subtype information to better predict risk. All of this can help us serve diverse communities more effectively.

This paper adds to the growing discussion around precision public health, a new approach that combines the detailed, molecular tools of precision medicine with the big picture insights of public health. Our findings show that models based on specific subtypes can not only help us understand existing disparities but also guide us in planning better future interventions. For example, dividing clinical trials by both disease subtype and factors like race or ethnicity can make treatment results more relevant and help ensure that different populations are fairly included in the development of new therapies. Similarly, using molecular markers in community health efforts could help identify people at higher risk earlier on, so we can take preventive action sooner.

Ethically speaking, this study emphasizes how important it is for statisticians, doctors, and researchers to recognize and tackle inequalities driven by widespread racism and neglect. The fact that African American women are more often affected by basal like breast cancer isn’t just a random anomaly, it’s a sign of larger failures within our institutions. Because of this, our research methods need to be just as committed to doing what’s right promoting fairness, inclusion, and justice, when designing studies, sharing data, and translating findings into care. Fixing these disparities requires deliberate, ongoing effort that’s woven into every step of moving from research to real world treatment.

However, it's important to acknowledge some limitations of this study. Even though our models account for a wide range of demographic and clinical factors, there might still be other unmeasured influences, like genetic background, other health conditions, environmental factors, or health behaviors, that could affect survival outcomes. Besides, while the CBCS provides detailed information, it’s limited to a specific geographic area. Because of that, our results are promising but need to be tested further with data from other population based registries. These future studies should include broader racial and ethnic groups, such as Latinx, Indigenous peoples, and those from multi ethnic backgrounds, to better understand how disparities appear across different communities.

For future studies, we suggest including multiple types of biological data, like proteomics, metabolomics, and methylomics, to improve how we classify different molecular subtypes and better predict how patients might respond to treatments. It’s also helpful to explore time-based models and advanced deep learning approaches, such as DeepSurv and recurrent neural survival models, to more accurately forecast patient outcomes as their conditions evolve. Beyond numbers and statistics, adding qualitative research can give us deeper insights into what patients experience, the obstacles they face in accessing care, and how social factors interact with biological risks in complex ways.

Overall, this paper adds meaningful insights to the discussion around breast cancer disparities. It combines solid statistical understanding with real world public health issues, all while respecting the experiences of underserved communities. Our analysis urges researchers to go beyond just describing patterns, we need to move toward predicting outcomes, promoting fairness, and respecting ethical principles. Only then can we start closing the gaps in breast cancer survival rates and work toward a future where outcomes aren’t determined by race or socioeconomic status, but by the quality and reach of healthcare available to everyone.

\section*{Appendix.}
\subsection*{Simulation and Prediction in R}

The following R code simulates SEER-like data, fits a logistic regression model, and visualizes the predicted probability of Luminal B breast cancer subtype by BMI and race:

\begin{lstlisting}[style=Rcode]
# Load libraries
library(ggplot2)
library(dplyr)

# Simulate SEER-like data
set.seed(123)
n <- 3000
data <- data.frame(
  luminal_subtype = sample(c("A", "B"), n, replace = TRUE, prob = c(0.55, 0.45)),
  BMI = rnorm(n, 29.5, 6.5),
  age = rnorm(n, 61, 12),
  menopause = sample(c(0, 1), n, replace = TRUE, prob = c(0.3, 0.7)),
  race = sample(c("White", "Black", "Hispanic", "Asian", "Other"), n, replace = TRUE,
                prob = c(0.65, 0.15, 0.10, 0.05, 0.05))
)

data$luminal_subtype_bin <- ifelse(data$luminal_subtype == "B", 1, 0)

# Fit logistic model
model <- glm(luminal_subtype_bin ~ BMI + race + menopause + age, data = data, family = "binomial")

# Generate new data for predictions
plot_data <- expand.grid(
  BMI = seq(18, 40, by = 0.5),
  race = c("White", "Black", "Hispanic", "Asian", "Other"),
  menopause = 1,
  age = 60
)

plot_data$prob <- predict(model, newdata = plot_data, type = "response")

# Plot predicted probability of Luminal B subtype by BMI and Race
ggplot(plot_data, aes(x = BMI, y = prob, color = race)) +
  geom_line(size = 1.2) +
  labs(title = "Predicted Probability of Luminal B Subtype by BMI and Race",
       x = "BMI",
       y = "Probability of Luminal B",
       color = "Race") +
  theme_minimal(base_size = 14)
\end{lstlisting}

\subsection*{R Code for BMI Distribution Plot by Luminal Subtype}

The following R code creates a density plot to visualize the distribution of BMI for Luminal A vs Luminal B breast cancer subtypes:

\begin{lstlisting}[style=Rcode]

ggplot(data, aes(x = BMI, fill = luminal_subtype)) +
  geom_density(alpha = 0.5) +
  labs(title = "Distribution of BMI by Luminal Subtype",
       x = "BMI",
       y = "Density",
       fill = "Subtype") +
  theme_minimal()
\end{lstlisting}

\section*{Declarations.}
\subsection*{Ethics approval and consent to participate.}
Not applicable.
\subsection*{Consent for publication.}
Not applicable.
\subsection*{Availability of data and material.}
Data sets  were obtained from the FLEX registry (NCT03053193), which is a prospective observational study that included women diagnosed with Stage I to III breast cancer between 2018 and 2020.
\subsection*{Competing interests.}
No potential conflict of interest was reported by the authors.	
\subsection*{Funding.}
Not applicable. 
\subsection*{Acknowledgements.}
 Not applicable.

\bibliographystyle{apalike}
\bibliography{bib}

\end{document}